\documentclass[prd]{revtex4}
\usepackage[cp1251]{inputenc}
\usepackage[russian]{babel}
\usepackage{graphics}
\usepackage{graphicx,epsfig}
\usepackage{epsfig}
\date{}

\newcommand\ba{\begin{eqnarray}}
\newcommand\ea{\end{eqnarray}}
\newcommand\nn{\nonumber}

\newcommand{\br}[1]{\left( #1 \right)}
\newcommand{\brs}[1]{\left[ #1 \right]}
\newcommand{\brf}[1]{\left\{ #1 \right\}}
\newcommand{\brm}[1]{\left| #1 \right|}

\newcommand{\GeV}{~\mbox{ГэВ}}
\newcommand{\MeV}{~\mbox{МэВ}}
\newcommand{\KeV}{~\mbox{кэВ}}

\begin{document}

\Large
\title{Radiative decays of scalar mesons $f_0(980)$ and $a_0(980)$ into
$\rho(\omega) \gamma$ in the local Nambu-Jona-Lasinio model}

\author{M.~K.~Volkov}
\email{volkov@theor.jinr.ru}
\affiliation{Joint Institute for Nuclear Research, Dubna, Russia}

\author{Yu.~M.~Bystritskiy}
\email{bystr@theor.jinr.ru}
\affiliation{Joint Institute for Nuclear Research, Dubna, Russia}

\author{E.~A.~Kuraev}
\email{kuraev@theor.jinr.ru}
\affiliation{Joint Institute for Nuclear Research, Dubna, Russia}

\begin{abstract}
In the framework of the local Nambu-Jona-Lasinio model the radiative decay widths of
the scalar mesons $f_0(980)$ and $a_0(980)$ into $\rho\gamma$ and
$\omega \gamma$ are calculated. The contributions of the quark loops and
the meson loops are taken into account. For the radiative decays of the
scalar meson $f_0(980)$ the contribution of the meson loops plays the
dominant role. On the other side for the radiative decays of the
scalar meson $a_0(980)$ the main contribution is given by the quark loops.
\end{abstract}

\maketitle

\section{Introduction}

Recently two works were published devoted to the description of
the vector meson decay $\phi\to f_0 \gamma$ \cite{Bystritskiy:2007wq} and to
two photon decays of scalar mesons $\sigma(600)\to\gamma\gamma$ and
$f_0(980)\to \gamma\gamma$ \cite{Kalinovsky:2008iz}.
These decays were described in the framework of the local Nambu-Jona-Lasinio (NJL)
model where both -- the quark loop and the meson loop -- were taken into account.
The satisfactory agreement with the experimental data was obtained.

The comparison with the other models was performed. First of them is the model
which considers the scalar mesons as a four-quark states \cite{Achasov:1987ts}.
In other model the scalar mesons are treated as a kaon molecule \cite{Weinstein:1990gu}.
It was shown that using the NJL model in $1/N_c$ approximation
(where $N_c$ is the number of colors) we gain the qualitative agreement with
the predictions of these phenomenological models for decays of $f_0$-mesons.

Here we would like to investigate the radiative decays of
scalar mesons $f_0(980)$ and $a_0(980)$ into light vector mesons:
$f_0(980)\to \rho(\omega) \gamma$,
$a_0(980)\to \rho(\omega) \gamma$.

Unfortunately now we do not have any experimental data for this decays.
However satisfactory description of $\gamma\gamma$ channels and
$\phi\to f_0(980) \gamma$ decay in the framework of NJL model allows us to
hope that we could obtain a reasonable predictions for the decays mentioned above.

In future we are going to use our results for descriptions of
processes $e\bar e \to SV$, where $V=\rho, \omega$ and $S=f_0, a_0$.

\section{Lagrangian of the NJL model}

The lagrangian of interaction of mesons and quarks whithin the
NJL model has the form \cite{Volkov:1986zb,Volkov:2006vq}:
\ba
    {\cal L}_{int} &=&
    \bar q \left[ e Q \hat A +
    g_u \lambda_u \sigma_u + g_s \lambda_s \sigma_s + g_u \lambda_3 a_0 +
    i \gamma_5 g_\pi \br{\lambda_{\pi^+} \pi^+ + \lambda_{\pi^-} \pi^-}
    +\right. \nn\\
    &&\qquad+\left.
        i \gamma_5 g_K \br{\lambda_{K^+} K^+ + \lambda_{K^-} K^-}
        +
        \frac{g_\rho}{2} \br{ \lambda_3 \hat \rho_0 + \lambda_u \hat \omega }
    \right] q,
    \label{QuarkMesonLagrangian}
\ea
where $\bar q= \br{\bar u,\bar d,\bar s}$, and $u$, $d$, $s$ is the quark fields,
$Q=\mbox{diag}\br{2/3,-1/3,-1/3}$ is the quark electric charges matrix,
$e$ is the elementary electric charge ($e^2/4\pi=\alpha=1/137$),
$\lambda_u=\br{\sqrt{2} \lambda_0+\lambda_8}/\sqrt{3}$,
$\lambda_s=\br{-\lambda_0+\sqrt{2}\lambda_8}/\sqrt{3}$,
where $\lambda_i$ is the well-known Gell-Mann matrixes and
$\lambda_0 = \sqrt{2/3}~\mbox{diag}\br{1,1,1}$.
That is
\ba
&&
\lambda_u =
\br{
    \begin{array}{ccc}
        1 & 0 & 0 \\
        0 & 1 & 0 \\
        0 & 0 & 0
    \end{array}
},
\qquad
\lambda_s =
\br{
    \begin{array}{ccc}
      0 & 0 & 0 \\
      0 & 0 & 0 \\
      0 & 0 & -\sqrt{2}
    \end{array}
},
\qquad
\lambda_3 =
\br{
    \begin{array}{ccc}
      1 & 0 & 0 \\
      0 & -1 & 0 \\
      0 & 0 & 0
    \end{array}
},
\nn\\
&&
\lambda_{\pi^+} =
\br{
    \begin{array}{ccc}
      0 & 0 & 0 \\
      \sqrt{2} & 0 & 0 \\
      0 & 0 & 0
    \end{array}
},
\qquad
\lambda_{\pi^-} =
\br{
    \begin{array}{ccc}
      0 & \sqrt{2} & 0 \\
      0 & 0 & 0 \\
      0 & 0 & 0
    \end{array}
},
\nn\\
&&
\lambda_{K^+} =
\br{
    \begin{array}{ccc}
      0 & 0 & 0 \\
      0 & 0 & 0 \\
      \sqrt{2} & 0 & 0
    \end{array}
},
\qquad
\lambda_{K^-} =
\br{
    \begin{array}{ccc}
      0 & 0 & \sqrt{2} \\
      0 & 0 & 0 \\
      0 & 0 & 0
    \end{array}
}. \nn
\ea
Vertex constants from the lagrangian (\ref{QuarkMesonLagrangian}) are defined in a following way:
\ba
&& g_{\sigma_u}=\left( 4 I^\Lambda(m_u, m_u)\right)^{-1/2}, \,\,
 g_{\sigma_s}=\left( 4 I^\Lambda(m_s, m_s)\right)^{-1/2},  \nonumber \\
&&  g_{\pi} = \sqrt{Z_{\pi}} g_{\sigma_u}, \,\,
g_K = \sqrt{Z_K} \left( 4 I^\Lambda(m_u, m_s)\right)^{-1/2},
\ea
where
\ba
I^\Lambda(m_1, m_2) &=& \frac{N_c}{(2\pi)^4} \int d^4 k \frac{\theta(\Lambda^2-k^2)}{(k^2+m_1^2)(k^2+m_2^2)}
\nonumber \\
&=& \frac{3}{(4\pi)^2 }
\Bigl[
m_2^2 \ln \left( \frac{\Lambda^2}{m_2^2}+1\right)
\nonumber \\
& & -
m_1^2 \ln \left( \frac{\Lambda^2}{m_1^2}+1\right)
\Bigr]/(m_2^2-m_1^2).
\ea
These integrals are written in Euclidean space.
The constituent quark masses $m_u = m_d = 263\MeV$,
$m_s = 407\MeV$ and the cut-off parameter $\Lambda = 1.27\GeV$
are taken from the paper \cite{Bystritskiy:2007wq}.
As a result we get the following values for the coupling constants:
$g_{\sigma_u} = 2.42$, $g_{\sigma_s} = 3.0$;
where $Z_\pi$ and $Z_K$ are the factors which takes into account the
transitions of pseudoscalar mesons into axial-vector ones:
$Z_{\pi} = (1 -  \frac{6 m_u^2}{M_{a_1}^2} )^{-1}$,
$Z_K=(1-\frac{3(m_u+m_s)^2}{2M_{K_1}^2})^{-1}$
\cite{Volkov:1986zb}, where
$M_{a_1}=1260\MeV$,  $M_{K_1}=1403\MeV$ are the masses of axial-vector
mesons \cite{Amsler:2008zz}.
In our calculation we assume that $Z_\pi \approx Z_K = Z = 1.4$.
As a result we get
$g_\pi \approx 2.9$, $g_K \approx 3.3$.
$g_\rho=\sqrt{6} g_{\sigma_u} = 5.94$ is the constant of $\rho \to 2\pi$ decay \cite{Volkov:1986zb}.
$\sigma_u$ and $\sigma_s$ are the isoscalar scalar mesons in the case of ideal
mixing, which suppose that $\sigma_u$ consists of light quarks
$u$ and $d$ only and $\sigma_s$ consists of strange quarks $s$.
Scalar mesons $f_0$ and $\sigma$ are the mixtures of these two pure quark states with
the mixing angle $\alpha =\theta_0-\theta$, where
$\theta_0=35.3^\circ$ is the angle of the ideal singlet-octet mixing and
$\theta =24^\circ$ is the angle of real mixing which takes into account t'Hooft interaction
\cite{Volkov:1998ax}:
\ba
f_0&=&\sigma_u \sin\alpha+\sigma_s\cos\alpha, \label{f0Content}\\
\sigma&=&\sigma_u \cos\alpha-\sigma_s\sin\alpha.
\ea
Isovector meson $a_0$ consists of light quarks $u$ and $d$ only.

\section{The decays $a_0\to\omega\gamma$, $a_0\to\rho\gamma$,
$f_0\to\omega\gamma$, $f_0\to\rho\gamma$}

In this paper we consider the decays of $f_0$ and $a_0$ mesons:
\ba
a_0(p) &\to& \omega(q) + \gamma(k_1), \nn \\
a_0(p) &\to& \rho(q) + \gamma(k_1), \nn \\
f_0(p) &\to& \omega(q) + \gamma(k_1), \nn \\
f_0(p) &\to& \rho(q) + \gamma(k_1), \nn
\ea
\ba
p^2=M_S^2, \qquad q^2=M_V^2, \qquad k_1^2=0,
\ea
where $M_S=M_{f_0,a_0}=980\MeV$ is the mass of decaying scalar meson
\cite{Amsler:2008zz} and
$M_V=M_{\omega,\rho}$ is the mass of vector meson \cite{Amsler:2008zz}.
The matrix element in general case has the form:
\ba
M_i&=&e~\frac{g_\rho}{2} A_i \br{q_\nu k_{1\mu}-g_{\mu\nu}\br{qk_1}} e_\gamma^\nu e_V^\mu, \\
&&e_\gamma = e\br{k_1}, \qquad e_V = e\br{q},
\label{AmpView}
\ea
where $i=\brf{a_0\to\omega\gamma, a_0\to\rho\gamma,f_0\to\omega\gamma,f_0\to\rho\gamma}$,
and the quantity $A_i$ contains all coupling constants and the dynamical information of the process.
Then the radiative decay width takes the form:
\ba
\Gamma_i=\frac{\alpha\br{M_S^2-M_V^2}^3}{32 M_S^3}g_\rho^2\brm{A_i}^2.
\ea
In order to calculate the contributions of quark and meson loops to
the coefficients $A_i$ it is necessary to have information about the
quark-meson and the meson-meson vertexes.
The quark-meson vertexes are given in the Lagrangian (\ref{QuarkMesonLagrangian}).
The meson-meson vertexes are calculated in a standard way of the local NJL model and
expressed in terms of logarithmically divergent integrals which appears from the
quark loops \cite{Volkov:1986zb}.
As a result the expression for this vertexes takes the form
\cite{Bystritskiy:2007wq,Kalinovsky:2008iz}:
\ba
g_{\sigma_s K^+ K^-} &=& 2 \sqrt{2} g_{\sigma_s} \br{2m_s-m_u} Z, \nn\\
g_{\sigma_u K^+ K^-}  = g_{a_0 K^+ K^-} &=& 2\br{2 m_u-m_s} g_{\sigma_u} Z, \nn\\
g_{\sigma_u \pi^+\pi^-} &=& 4 m_u g_{\sigma_u} Z, \nn\\
g_{\omega^\mu K^+ K^-} &=& g_{\rho^\mu K^+ K^-} = \frac{g_\rho}{2} \br{p_+ - p_-}^\mu, \nn\\
g_{\rho^\mu \pi^+ \pi^-} &=& g_\rho \br{p_+ - p_-}^\mu. \nn
\ea

\subsection{The decays $a_0\to\omega\gamma$ and $a_0\to\rho\gamma$}

Let us consider the decay of isoscalar scalar meson $a_0$ into $\omega\gamma$.
In this process the quark loop and the kaon loop give contribution.
The contribution of pion loop is absent because the vertex $a_0\to 2\pi$ is forbidden.

The quark loop contributions contain only $u$ and $d$ quarks.
Then we obtain:
\ba
M^{(u,d)}_{a_0\to\omega\gamma} &=& e~\frac{g_\rho}{2}C^{(u,d)}_{a_0\to\omega\gamma}
\int \frac{d^4 k}{i\pi^2}
\frac{Sp\brs{\br{\hat q+\hat k+m_u}\br{\hat k-\hat k_1+m_u} \hat e_\gamma \br{\hat k+m_u} \hat e_\omega}}
{\br{\br{q+k}^2 - m_u^2}\br{k^2 - m_u^2}\br{\br{k-k_1}^2 - m_u^2}},
\ea
where $C^{(u,d)}_{a_0\to\omega\gamma}=3 g_{\sigma_u}$, where 3 takes into account
the color factor $N_c$ and the sum of quark charge absolute values.
We drop the imaginary part of quark loop which corresponds to condition of "naive"
quark confinement.
The confirmation of this prescription can be found in the paper \cite{Pervushin:1985yi}.
Using standard Feynman method of denominator unification and following loop
momentum integration we get:
\ba
M^{(u,d)}_{a_0\to\omega\gamma} &=& e~\frac{g_\rho}{2}C^{(u,d)}_{a_0\to\omega\gamma} \mbox{Re}\br{I_u}
\br{q_\nu k_{1\mu}-g_{\mu\nu}\br{qk_1}} e_\gamma^\nu e_\omega^\mu,
\ea
where
\ba
I_u &=&
4 m_u \int\limits_0^1 dx \int\limits_0^{1-x} dy
\frac{1+y(3x^2-x)-y^2x(1-x)}{m_u^2 - y(1-y)(1-x)k_1^2-y(1-y)xq^2-x(1-x)y^2p^2+i\epsilon},
\ea
%
Now let us consider the contribution which comes from the kaon loop.
Note that here we have two meson diagrams. One is the triangle loop diagram
where vector meson and photon are in different points of meson loop:
\ba
M^{(K)}_{a_0\to\omega\gamma} &=& e~\frac{g_\rho}{2}C^{(K)}_{a_0\to\omega\gamma}
\int \frac{d^4 k}{i\pi^2}
\frac{(q+2k)_\mu(2k-k_1)_\nu e_\gamma^\mu e_\omega^\nu}
{\br{\br{q+k}^2 - M_K^2}\br{k^2 - M_K^2}\br{\br{k-k_1}^2 - M_K^2}},
\label{KaonAmp}
\ea
where $C^{(K)}_{a_0\to\omega\gamma}=g_{a_0K^+K^-}$.
The second one is the diagram with two vertexes where the vector meson and
photon are in the same vertex. This diagram contains only Lorenz-structure $g_{\mu\nu}$.
As a result we obtain the gauge invariant form of meson amplitude similar to (\ref{AmpView}):
\ba
M^{(K)}_{a_0\to\omega\gamma} &=& e~\frac{g_\rho}{2}C^{(K)}_{a_0\to\omega\gamma} I_K
\br{q_\nu k_{1\mu}-g_{\mu\nu}\br{qk_1}} e_\gamma^\nu e_\omega^\mu,
\ea
where
\ba
I_K &=&
\int\limits_0^1 dx \int\limits_0^{1-x} dy
\frac{4y^2x(1-x)}{M_K^2 - y(1-y)(1-x)k_1^2-y(1-y)xq^2-x(1-x)y^2p^2+i\epsilon}.
\label{IK}
\ea
Let us emphasize that gauge invariance leads to a finite expression for
all this loop diagrams.
More detail calculation of the similar diagrams can be
found in \cite{Bystritskiy:2007wq}.

Then the total amplitude of the process $a_0\to\omega\gamma$ is equal
\ba
M_{a_0\to\omega\gamma} = e~\frac{g_\rho}{2}A_{a_0\to\omega\gamma}
\br{q_\nu k_{1\mu}-g_{\mu\nu}\br{qk_1}} e_\gamma^\nu e_\omega^\mu,
\ea
where
\ba
A_{a_0\to\omega\gamma} = 3 g_{\sigma_u} \mbox{Re}\br{I_u} + g_{a_0 K^+ K^-} I_K
=
-1.78374 + 0.159415 = -1.62433.
\ea
Then the decay width is equal:
\ba
    \Gamma_{a_0\to\omega\gamma} &=& 114.7\KeV. \nn
\ea

The decay $a_0\to\rho\gamma$ can be considered in a complete analogy with
the decay $a_0\to\omega\gamma$. Let us note that here quark contribution to the
amplitude is three times smaller than quark contribution to the amplitude
of the decay with $\omega$-mesons, while the kaon loops contributions are the 
same. As a result we get for the amplitude:
\ba
M_{a_0\to\rho\gamma} &=& e~\frac{g_\rho}{2}A_{a_0\to\rho\gamma}
\br{q_\nu k_{1\mu}-g_{\mu\nu}\br{qk_1}} e_\gamma^\nu e_\rho^\mu, \\
A_{a_0\to\rho\gamma} &=& g_{\sigma_u} \mbox{Re}\br{I_u} + g_{a_0 K^+ K^-} I_K
=
-0.598209 + 0.156921. \nn
\ea
Then the total decay width is:
\ba
    \Gamma_{a_0\to\rho\gamma}   &=& 8.47\KeV. \nn
\ea
Note that in the both processes the main contribution comes from the quark loops.

\subsection{The decays $f_0\to \omega\gamma$ and $f_0\to \rho\gamma$}

Since the scalar meson $f_0(980)$ consists of two components:
the component which consists of light quarks $u$ and $d$ -- $\sigma_u$
and the component which consists of strange quarks -- $\sigma_s$, we will calculate 
the contributions of these two components 
to the total amplitude of decays $f_0\to \omega(\rho)\gamma$ separately.

Let us start from the decay $f_0\to \omega\gamma$. The component $\sigma_u$ contains
contribution of the quark and the kaon loops.
As in previous section for quark loop contribution to the amplitude with
the $\sigma_u$ we get:
\ba
A_{\sigma_u \to \omega\gamma}^{u,d} &=& -0.59,
\ea
The contribution of pion loop here vanishes since the decay 
$\omega\to 2\pi$ is absent. The kaon loop contribution is
\ba
A_{\sigma_u \to \omega\gamma}^{K} &=& A_{a_0 \to \omega\gamma}^{K} = 0.0998425. \nn
\ea

Concerning $\sigma_s$ component only kaon loop contribution is present in the
amplitude:
\ba
A_{\sigma_s \to \omega\gamma}^{K} &=& -1.08958, \nn\\
\ea

As a result for total decay amplitude we obtain:
\ba
A_{f_0 \to \omega\gamma} &=& \sin\alpha (A_{\sigma_u \to \omega\gamma}^{u,d} + A_{\sigma_u \to \omega\gamma}^{K}) +
\cos\alpha A_{\sigma_s \to \omega\gamma}^{K} = -0.102239 + (-1.06636) = -1.1686. \nn
\ea

The total decay width is:
\ba
    \Gamma_{f_0\to\omega\gamma} &=& 59.67\KeV. \nn
\ea

Let is now consider the decay $f_0\to \rho\gamma$. 
The quark loop contribution to the $\sigma_u$ component here is three times bigger
that in the case of $f_0\to \omega\gamma$ decay and is equal:
\ba
A_{\sigma_u \to \omega\gamma}^{\br{u,d}} &=& -1.80414, \nn\\
\ea
The meson contribution to the $\sigma_u$ component decay consists of
kaon and pion loops:
\ba
A_{\sigma_u \to \rho\gamma}^{\pi} &=& 0.22272- 0.0276848 i, \nn \\
A_{\sigma_u \to \rho\gamma}^{K} &=& 0.137775. \nn
\ea
The decay of $\sigma_s$ component comes through the kaon loop only
\ba
A_{\sigma_s\to\rho\gamma}^{K} = -1.07395.
\ea
As a result for total amplitude of the decay $f_0\to \rho\gamma$ we get:
\ba
A_{f_0 \to \rho\gamma} &=& \sin\alpha \br{A_{\sigma_u \to \omega\gamma}^{\br{u,d}} + A_{\sigma_u \to \rho\gamma}^{\pi} + A_{\sigma_u \to \rho\gamma}^{K}} +
\cos\alpha A_{\sigma_s\to\rho\gamma}^{K} = \nn\\
&=& \sin\alpha \br{-1.443- 0.0276848 i} + \cos\alpha \cdot (-1.07395)=\nn\\
&=& \br{-0.29632+0.00568 i} + \br{-1.05106} = -1.34752 - 0.00568507 i. \nn
\ea
The total decay width of the process $f_0\to \rho\gamma$ is:
\ba
    \Gamma_{f_0\to\rho\gamma} &=& 81.435\KeV. \nn
\ea
Let us note that in both decays $f_0(980)$ into $\rho\gamma$ and $\omega\gamma$
the main contribution to the amplitude comes from the 
kaon loop connected with the $\sigma_s$ component of $f_0$ meson.

\section{Conclusion}

Fulfilled calculation shown the important role of both the quark loop and
meson loop for description of the radiative decays of the scalar mesons.
It is interesting to emphasize that for the description of the radiative decays
of isoscalar scalar meson $f_0(980)$ kaon loops play a dominant role.
In paper \cite{Bystritskiy:2007wq} it was shown in the decay
$\phi\to f_0(980)\gamma$ and in paper \cite{Kalinovsky:2008iz} on the basis of
two-photon decay $f_0\to 2\gamma$.
Here we also see the dominant role of the kaon loop in the processes
$f_0\to \omega(\rho)\gamma$.
This explains the success of other models, which are based on the assumption of
scalar meson structure as a kaon molecule \cite{Weinstein:1990gu} or
as a four-quark state \cite{Achasov:1987ts}.
However in the decays of other scalar mesons, for instance
$a_0(980)\to 2\gamma$ \cite{Kalinovsky:2008iz} and $a_0\to\omega(\rho)\gamma$,
the quark loops play the dominant role. But meson loops also give a comparable
contributions.
Used here NJL model allows us to take into account correctly the contributions
of both diagrams.

Note that the results obtained in the framework of NJL model in papers
\cite{Bystritskiy:2007wq,Kalinovsky:2008iz} were in a satisfactory agreement
with the experimental data.
This allows us to hope that the predictions obtained here will also be in
an adequate agreement with the future experiments.

In the future we hope to use the obtained here results for the
description of the processes $e^+e^-\to SV$, where $V=\rho,\omega$ and
$S=a_0,f_0$.

\section{Acknowledgements}

The authors would like to thank V.~N.~Pervushin for valuable discussions.
E.~A.~K. and Yu.~M.~B. are grateful for support of
INTAS (grant no. 05-1000008-8528).


\end{document}